\documentclass[onecolumn]{aastex6}

\newcounter{magicrownumbers}

\usepackage{lineno}

\slugcomment{The Astronomical Journal, in press - this is the ArXiv copy}
\shorttitle{UV Spectroscopy of T Aurigae
}
\shortauthors{  }
\setwatermarkfontsize{1.1in}

\begin{document}

\title{{\bf Hubble Space Telescope STIS Spectroscopy of Nova T Aurigae 1891}}

\author{Conor Larsen\altaffilmark{1} }
\author{Patrick Godon\altaffilmark{1,2}}
\author{Edward M. Sion\altaffilmark{1} }


\email{clarsen2@villanova.edu} 
\email{patrick.godon@villanova.edu}
\email{edward.sion@villanova.edu}

\altaffiltext{1}{Department of Astrophysics \& Planetary Science, 
Villanova University, Villanova, PA 19085, USA}
\altaffiltext{2}{Henry A. Rowland Department of Physics \& Astronomy,
The Johns Hopkins University, Baltimore, MD 21218, USA}

\begin{abstract}
T Aurigae is an eclipsing old nova which exploded in 1891. 
At a Gaia EDR3 distance of 815-871 pc, it is a relatively nearby old nova. 
Through ultraviolet spectral modeling and using the new precise Gaia distance, 
we find that the HST/STIS spectrum of T Aurigae is consistent with an accretion disk 
with a mass transfer rate $\dot{M}$ of the order of $10^{-8}M_{\odot}$/yr, 
for a white dwarf mass of $M_{\rm wd} \approx 0.7 \pm 0.2 M_{\odot}$, an 
inclination of $i \sim 60^{\circ}$, and a Gaia distance of of $840_{-25}^{+31}$~pc. 
The sharp absorption lines of metals cannot form in the disk and are likely forming in
material above the disk (e.g. due stream disk overflow), in circumbinary 
material, and/or in material associated with the ejected shell from the 1891 nova explosion.
The saturated hydrogen Ly$\alpha$ absorption feature is attributed to a large interstellar medium   
hydrogen column density of the order of $10^{21}$cm$^{-2}$ towards T Aur, 
as corroborated by the value of its reddening $E(B-V)=0.42 \pm 0.08$.  
\\

\end{abstract}


\section{Introduction} \label{sec:intro}



T Aurigae (henceforth T Aur) is a cataclysmic variable consisting of a white dwarf in a compact
 binary with a Roche lobe-filling main sequence-like donor star. The system underwent a classical nova explosion in 
 1891. As hydrogen-rich 
 material builds up on the surface of the 
 white dwarf, the critical pressure and ignition temperature is reached at the base of the accreted envelope 
 for hydrogen fusion in electron degenerate layers. Hence, nuclear energy is released but there is no compensating
 expansion of the burning region to cool 
 because the equation of state of the degenerate matter is insensitive to temperature. A runaway 
 thermonuclear explosion occurs identified as the classical nova \citep{mes52,sta71}.   
The evolution of old novae is poorly understood for several reasons. Most studies focus on  either, the nova explosion 
itself, the decline from outburst and abundances of metals from the emission lines that form in the nova shell as it expands.
Moreover, the time baseline of even the oldest old nova is too short to determine the long term accretion rate 

\par T Aur is a relatively nearby nova. The Gaia EDR3 parallax restricts the distance to
$d=840_{-25}^{+31}$~pc, or $d \approx 815 - 871$~pc, 
where we follow \citet{sch18} to derive the distance using the Gaia EDR3 parallax 
and error (note that in \citet{sch18} the Gaia DR2 data gives $d=880_{-35}^{+46}$~pc). 
\cite{walker} discovered that T Aur is an eclipsing binary with a period of
 4 hours and 54 minutes. The light curve is of Algol nature and no secondary eclipse is observed.
\cite{walker} also noted smaller variations in the light output (flickering) which are present in the current
 AAVSO light curves. These smaller light variations are also present in the nova system DQ Her
 and are indicative of  accretion (Hege et al.1980). Following the nova in 1891, T Aur brightened significantly,
 reaching a visual magnitude of 4.5. Since the explosion, T Aur has returned to quiescence, settling
 at a visual magnitude of 14.9 \citep{strope}. 

 \par Using the classification developed by \cite{strope}, the light curve of T Aur 1891 eruption is a D type,
 representing a dust dip type. Following an increase in brightness from the nova outburst, T Aur
 underwent a fast decline before reaching a minimum and returning to quiescence. This large dip
 in brightness is almost certainly caused by the formation of dust particles in the expanding nova 
 shell as the temperature of the shell decreases. As the dust
 particles continue to expand and attenuate, the brightness recovers. T Aur also has an impressive
 nova shell which has been imaged and studied by \cite{santamaria}. The shell is moving rapidly
 outward with an expansion velocity along the minor axis of 315 km/s.
System parameters have been previously derived and given in Table 1  as 
 reported in the literature.

Most of the previous analyses \citep[e.g.][]{pat84,dub18} were completed using optical
 observations, which may be affected by the stream impact zone at the
 edge of the accretion disk, the Roche lobe-filling secondary donor star or surrounding nebular material.

Our analysis utilizes
observations in the far ultraviolet wavelength range where the spectral energy distribution of 
the accretion disks and the white dwarf photospheres have their peak intensities. This minimizes or 
eliminates some of the contamination of the UV spectrum due to the hot spot (impact zone), the secondary donor star, 
and/or surrounding nebular material.

\par In section 2, we describe the synthetic spectral model codes for disks and photospheres that 
we employ, in section 3, we present the model fitting analysis of the HST spectrum and in section 4, 
we discuss our results and summarize our conclusions.

\begin{deluxetable*}{lcl} 
\tablewidth{0pt}
\tablecaption{Orbital and Physical Parameters of T Aurigae}
\tablehead{ 
Parameter        & Value        & References        
}
\startdata
Period & 0.204 days & \citet{walker} \\ 
Inclination & $57 \pm 8 ^{\circ}$ & \cite{sel19} \\ 
Distance & $840_{-25}^{+31}$~pc & Gaia EDR3 \\
$E(B-V)$ & 0.42$\pm$0.08 & \cite{sel2013} \\ 
Secondary Type & K5 & \cite{bianchini}\\ 
WD Mass & 0.7$\pm$0.2 $M_{\odot}$ & \cite{sel19} \\ 
WD Radius (quiescence) & 0.011 $R_{\odot}$ & \cite{bianchini}\\ 
Secondary Mass & 0.63 $M_{\odot}$ & \cite{bianchini} \\ 
Secondary Temperature  & $\sim$ 4000~K &           \\ 
\enddata
\end{deluxetable*}

\clearpage 

\section{Archival Data}

\begin{figure}[h!] 
\vspace{-3.cm} 
\epsscale{0.8} 
\plotone{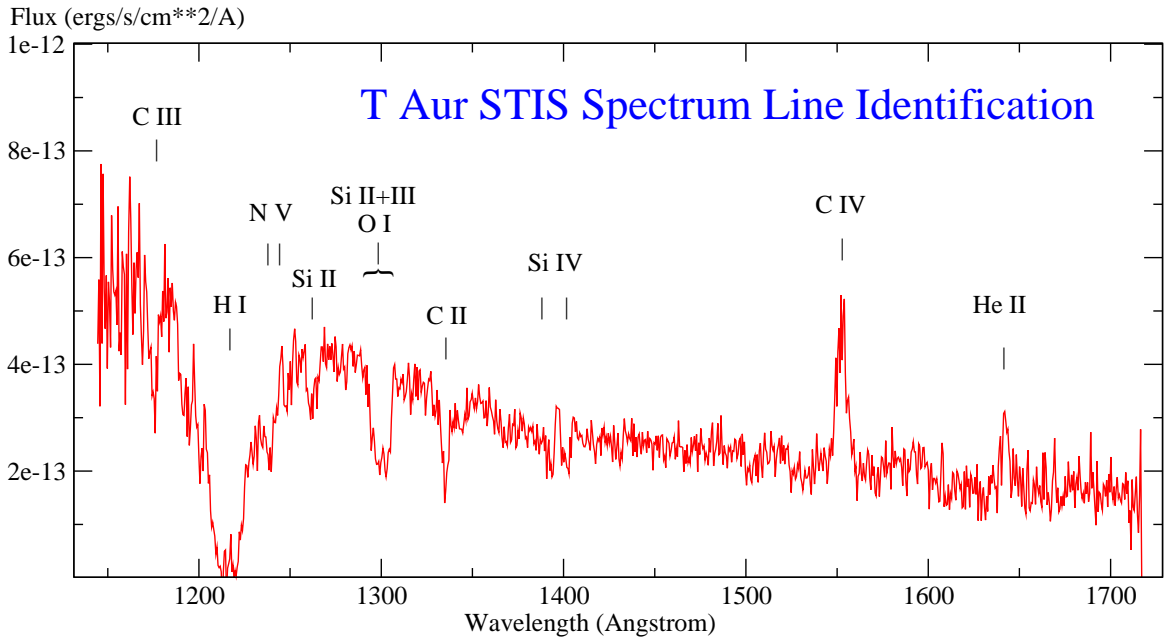} 
\caption{ 
The HST/STIS spectrum of the Nova T Aurigae 1891 is shown with line identification. 
The spectrum has been dereddened assuming $E(B-V)=0.42$. 
\label{spectrum}
}
\end{figure}

We retrieved the Hubble Space Telescope Imaging Spectrograph 
data from the MAST archive that had been obtained during a quiescent
interval on March 7, 2003 (UT start time 12:01:54). The data (O6LI0M010) were collected in ACCUM mode 
using the 52X0.2 aperture, and totaled 600~s of good exposure time. The FUV-MAMA detector was set 
with the G140L grating centered at 1425~\AA , thereby covering the wavelength range 
$\sim$1150 - 1715 {\AA}. 
The data were processed through the pipeline with CALSTIS version 3.4.  
The extracted HST STIS spectrum is displayed in Fig.\ref{spectrum}, 
where it has been dereddened assuming $E(B-V)=0.42\pm0.08$ \citep{sel2013}.  
This value of the reddening agrees within the error bars 
with the value of $E(B-V)=0.356 \pm 0.081$  
assessed from the 3D map of the local ISM \citep{lal14,cap17,lal19}.
Since T Aur has a nebula, it is likely
that its nebular material contributes additional reddening to that of the ISM. 
In Fig.\ref{spectrum}, we identify absorption lines due to 
C\,{\sc iii} (1175), 
N\,{\sc v} doublet ($\sim 1240$),  
Si\,{\sc ii} (1260, 1265), 
C\,{\sc ii} (1335),  
Si\,{\sc ii+iii} ($\sim 1300$) + O\,{\sc i} ($\sim 1300$), 
Si\,{\sc iv} doublet ($\sim1393, 1402$), and the unidentified feature at 1500 {\AA}, first 
noted by Selvelli and Gilmozzi (2013; see their Table 14) who observed the same feature being present in the 
HST spectrum of the old nova DI Lac. Only two emission lines are evident: the C\,{\sc iv} (1550) and
the He\,{\sc ii} (1640) lines. At the time of the observation the altitude of the sun was $20^{\circ}$ 
below the Earth's limb, therefore minimizing airglow (Ly$\alpha$ and O\,{\sc i}) emissions.  

\clearpage

\section{Spectral Analysis and Results}

\subsection{A White Dwarf Scenario?} 

Our first glance at the spectrum gave us the impression that it resembled the FUV spectrum of a 
bare, hot accreting white dwarf with little evidence of an accretion disk. This initial impression 
was based upon the fact that no optically thick steady state accretion disk spectrum with 
solar composition at any inclination can replicate the rich absorption line spectrum of T Aur. 
The absorption lines in accretion disks at any inclination, except face-on, are Keplerian-broadened 
to the extent that the model absorption features are "washed out" far 
too shallow and broad. Furthermore, we noted the 
presence of the Si\,{\sc ii} 1265 {\AA} line which arises from an excited level, only 
populated at higher densities, typically the earmark of a white dwarf photosphere. Finally, the hydrogen Ly$\alpha$ absorption 
appears to be saturated which is still another earmark of an exposed accreting white dwarf. These 
characteristics led us to test the white dwarf interpretation with the following simple exercise 
before we carried out detailed modeling.

\par  We estimate the observed UV luminosity of T Aur by the mean UV 
flux level of the STIS spectrum over the spectral range of 700~\AA . 
We estimated the average flux from the STIS spectrum to be 
$3 \times 10^{-13}$ erg/cm$^2$/s/\AA. This average flux is multiplied 
by the wavelength range of 700~\AA\ to obtain $F_{\lambda}$. The observed luminosity is given by

\begin{equation}
L_{obs} = 4 \pi d^2 F_{\lambda}. 
\end{equation}
   
By using the Gaia distance of 840 pc, we obtain an observed luminosity on the order of $10^{34}$ erg/s. 
 
We compare this observed luminosity with the bolometric luminosity output of an accreting white dwarf at a 
temperature of 30,000~K and a radius of 0.012 $R_{\odot}$, using the Stefan-Boltzmann Law.  
The luminosity output of the accreting white dwarf is on the order of $10^{32}$ erg/s, two orders 
of magnitude smaller than the observed luminosity computed above.
 
If the UV flux was due to a white dwarf photosphere, the spectral analysis (see below) indicates that it 
would correspond to a WD with a gravity $log(g) \approx 8.0 $ and temperature
of 28,500~K. While such a gravity gives a WD mass of $\approx 0.65 M_{\odot}$, close to the 
$0.7 M_{\odot}$ of \citet{sel19}, the scaling of the spectrum gives a distance
of only 88~pc, or about 10 times shorter than the Gaia distance of 840~pc (which 
accounts for a flux 100 times smaller).    

For the luminosity of the accretion disk, we follow \citet{fra92}, and 
calculate the luminosity of the accretion disk by estimating how efficiently the rest mass energy,
$c^2$ per unit mass, of the accreted material is converted to radiation. This efficiency is 
parameterized by a dimensionless quantity, $\eta$, given by

\begin{equation}
\eta = \frac{ G M_{\rm wd}} {R_{\rm wd} c^2}.  
\end{equation} 
The bolometric luminosity of the accretion disk is then given by   
\begin{equation} 
L_{\rm disk} = \eta \dot{M} c^2. 
\end{equation}

Taking the same parameters we used for the white dwarf above, we find that the disk luminosity is 
of order of $10^{34}$~ergs/s for an accretion rate of 
$\sim 10^{-8}M_{\odot}$/yr. This disk luminosity is 
is of the same order of magnitude as the above derived observed UV luminosity.
Thus, accretion light from a disk fully accounts for the observed luminosity of T Aur.   

Our examination of the ISM in line of sight to T Aur enabled us to account for the saturation of the 
Ly$\alpha$ absorption. As shown in Fig. 4, the Ly$\alpha$ absorption line is accounted 
for by high column densities of H\,{\sc i} in the sight line to T Aur. 
This is consistent with the high value 
of E(B-V) = 0.42 listed in Table 1. 

The origin of the absorption lines, labelled 'stellar' by 
Selvelli \& Gilmozzi (2013; see their Table 14) is uncertain. We are confident that the absorption lines
do not form in the accretion disk nor in the white dwarf photosphere. We have modeled the 
absorption lines along the lines described in the section on the disk modeling. Unfortunately, we cannot
check for motions of the absorption features during the full exposure by examining subexposures,
since the spectrum is a single ``snapshot''.

\clearpage

\clearpage

\subsection{Disk Models} 

We have carried out accretion disk model fits to the STIS spectrum of T Aur. 
We use the grid of accretion disk models from \citet{wad98}. 
These disk models have solar composition, and were generated (using TLUSTY/SYNSPEC)
for discrete values of a CV system parameters. 
In the present work we used the models with WD mass $0.55 M_{\odot}$, $0.80 M_{\odot}$,
and $1.03 M_{\odot}$, combined with a binary inclination $i=41^{\circ}$ and $60^{\circ}$,
since the WD mass of T Aur is estimated to be $0.7 \pm 0.2 M_{\odot}$ and its
inclination is $57 \pm 8 ^{\circ}$. 
The mass transfer rates available range from 
$\dot{M}= 10^{-11} M_{\odot}$/yr to  
$\dot{M}= 10^{-8} M_{\odot}$/yr, in steps of 0.5 in $log( \dot{M} )$, 
intermediate values $\dot{M}$ are obtained by linear interpolation between
the available $\dot{M}$ values. 

\begin{figure}[b!] 
\epsscale{0.7}
\plotone{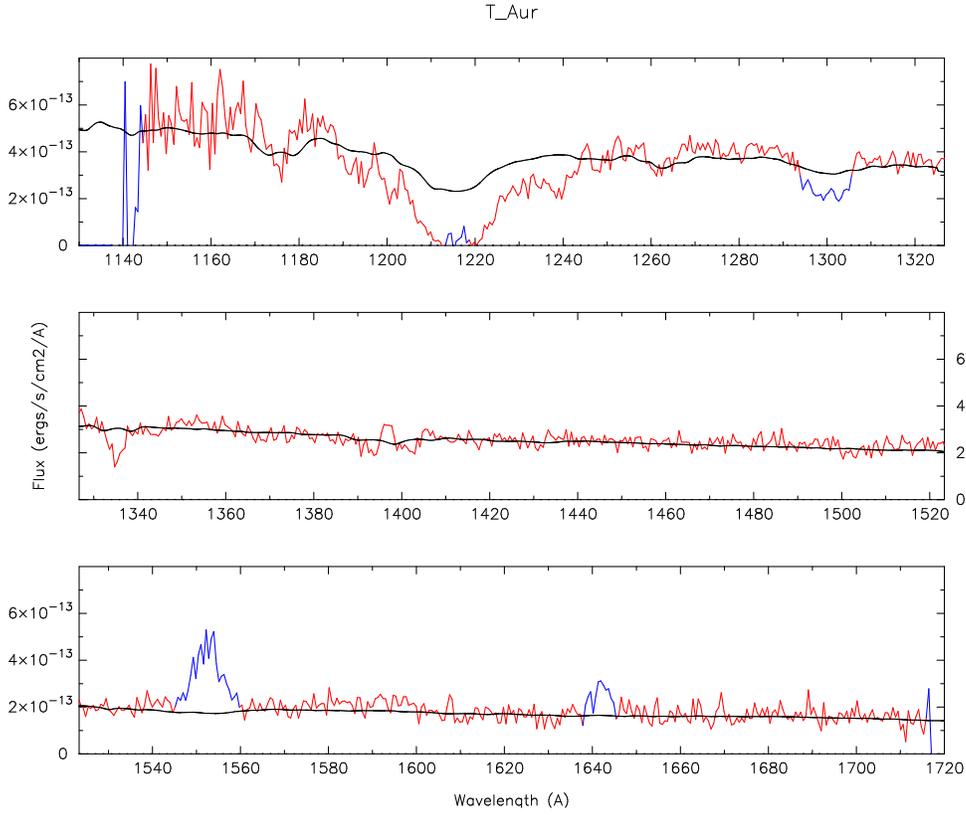}                 
\caption{
The spectrum of T Aur (in red) is fitted with an accretion
disk model (in black) with a WD mass $M_{\rm wd}=0.80M_{\odot}$,
an accretion rate $\dot{M}=10^{-8}M_{\odot}$/yr, and an
inclination $i=41^{\circ}$, giving a distance of 858~pc
and a $\chi^2_{\nu}$ value of 12.7. All the disk models
with $M_{\rm wd}$=0.55, 0.80, \& 1.03$M_{\odot}$ with 
an accretion rate of the order of $10^{-8}M_{\odot}$
and inclination $i=$41 \& 60$^{\circ}$ give about the 
same results (large $\chi^2_{\nu}$ and same distance)
as the Ly$\alpha$ region cannot properly be fitted with
such disk models. The regions in blue have been masked before
the fitting.  
\label{plaindisk} 
}
\end{figure}

Overall we find that disk models with the above WD masses 
(0.55, 0.80, and 1.03$M_{\odot}$) and inclinations ($41^{\circ}$ and $60^{\circ}$) 
scale to the correct Gaia distance for a mass accretion rate within a factor of
two of $\approx 10^{-8}M_{\odot}$/yr. However, all these models poorly fit
the Ly$\alpha$ region and the absorption lines.  In Fig.\ref{plaindisk} we present
such a  model fit with   
$M_{\rm wd}=0.8 M_{\odot}$, $\dot{M}=10^{-8}M_{\odot}$/yr, and $i=41^{\circ}$, 
scaling to a distance of 858~pc.

\clearpage 

Since the reddening towards T Aur is relatively large (compared to CVs in general,
but rather common for a nova at $\sim 1$~kpc), and that the system is enshrouded
by nebular material likely ejected during its 1891 nova eruption, we must consider that 
the Ly$\alpha$ region and absorption lines might be forming in material in the line
of sight of the observer. We therefore also model the interstellar medium (ISM) absorption and   
add absorbing slabs to the accretion disk models. 

The ISM modeling we use is described, e.g., in \citet{god07}. 
Assuming a reddening of $E(B-V) \approx 0.4$, one can assess the hydrogen column density  
to be of the order of $\approx 2 \times 20^{21}$cm$^{-2}$ \citep[see e.g.][]{boh78}. 
Though, because of the ejected nebular material, not all the reddening is due to 
the ISM. We find that an ISM modeling with a hydrogen column density $\sim 1 \times 10^{21}$cm$^{-2}$  
is enough to reproduce the Ly$\alpha$ absorption feature.  
The ISM model has a temperature of 50~K, a turbulent velocity of 80~km/s,  
and zero metallicity.  In the wavelength region considered here, 
the main ISM component affecting the spectrum is the atomic 
hydrogen which shapes the Ly$\alpha$ profile. The ISM molecular hydrogen column density was taken to 
be $\approx 10^{3}$ smaller than the atomic hydrogen column density and has no effect on the results 
\citep[the ISM H molecular lines appear at shorter wavelengths, see][for the molecular lines]{sem01}.   

On the other hand, 
we use SYNSPEC and Circus \citep{hub96} to generate the absorbing slabs and incorporate 
them into the disk models.  
Further details on the slab modeling can be found, e.g., in \citet[][]{hor94}.  
We first applied an iron curtain, i.e. a cold slab of material with a temperature of 10,000~K and an electron
density of $n_e=10^{13}$cm$^{-3}$. We increase the hydrogen column density, starting 
at $N_{\rm H}=10^{18}$cm$^{-2}$. We find that we are able to fit the silicon absorption features
at 1260 \& 1265~\AA\ together with the carbon line at 1335~\AA\ for 
$N_{\rm H}=5 \times 10^{18}$ with a turbulent dispersion velocity of 75~km/s. 
This also creates a nitrogen line at 1193~\AA .  
Since the remaining lines appear at a higher temperature, we applied a second
slab with T=20,000~K and find that for the same hydrogen column density and
turbulent velocity we are able to fit both the carbon line at 1175~\AA , and the
silicon resonance doublet at 1400~\AA . 
A third (similar) slab with a temperature of 40,000~K generates the nitrogen
doublet lines at 1240~\AA\ and a weak/shallow C\,{\sc iv} lines at 1550~\AA . 
All the absorbing slabs have solar composition. 
For completeness, we use the 3 slabs in the modeling together with the    
ISM absorption model.

%
%
%

With the help of the slabs and ISM modeling we are able to significantly improve the fits of
the disk models to the spectrum of T Aur. For example the disk model presented in Fig.\ref{plaindisk} 
has $\chi^2_{\nu} = 12.7$,  with the ISM modeling we are able to decrease it to $\chi^2_{\nu}=2.90$, 
and with the 3 absorbing slabs it comes down to $\chi^2_{\nu}=2.46$. 
We applied the ISM and absorbing slabs to all the disk models and present the results in 
Table \ref{diskmodels}.

\begin{deluxetable*}{cccccccc}[h!]  
\tablewidth{0pt}
\tablecaption{
Accretion Disk Models 
\label{diskmodels} 
}
\tablehead{ 
 $M_{\rm wd}$  &  $\dot{M}$       &   $i$   &   $d$   & $\chi^2_{\nu}$ & ISM  & Absorbing &  Fig. \\ 
 $(M_{\odot})$ & ($M_{\odot}$/yr) &  (deg)  &  (pc)   &         & Ly$\alpha$  & Slabs     &  \#    
}
\startdata
  0.55       & $1.5\times10^{-8}$ &  41     &  839    &  2.52  & Yes & Yes   &  ---    \\ 
  0.55       & $2.8\times10^{-8}$ &  60     &  833    &  2.63  & Yes & Yes   &  ---    \\ 
  0.80       & $1.0\times10^{-8}$ &  41     &  858    &  12.7  & No  & No    &  \ref{plaindisk}    \\ 
  0.80       & $1.0\times10^{-8}$ &  41     &  841    &  2.90  & Yes & No    &  ---    \\ 
  0.80       & $1.0\times10^{-8}$ &  41     &  842    &  2.46  & Yes & Yes   &  ---    \\ 
  0.80       & $1.8\times10^{-8}$ &  60     &  832    &  2.55  & Yes & Yes   &  ---    \\ 
  1.03       & $0.76\times10^{-8}$ &  41    &  839    &  2.30 & Yes & Yes   &  \ref{veileddisk}     \\ 
  1.03       & $1.37\times10^{-8}$ &  60     &  837    &  2.39 & Yes & Yes   &  ---     \\ 
\enddata
\tablecomments{All the disk models were assumed to have solar composition. 
}    
\end{deluxetable*}

\clearpage 

\begin{figure}[b!]  
\epsscale{0.7} 
\plotone{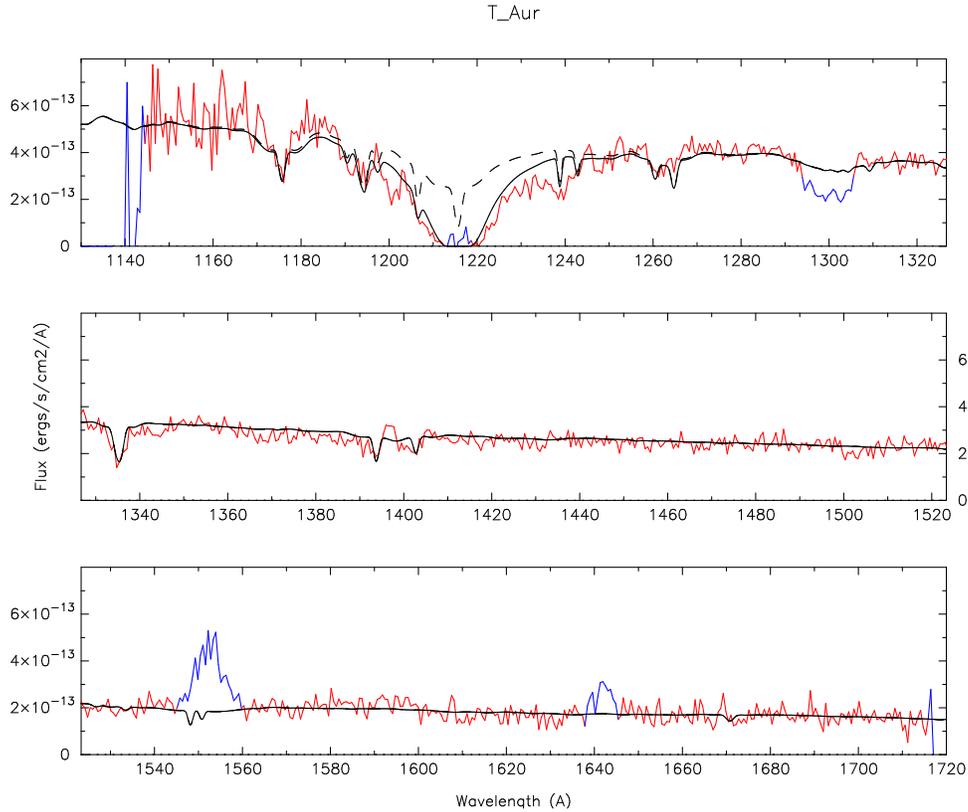}                 
\caption{The spectrum of T Aur (in red) is fitted with an accretion disk 
with  absorbing slabs and ISM absorption model (solid black line). 
The addition of 3 absorbing slabs  generates a spectrum with some absorption
lines (dashed black line); and the addition of the ISM models creates a 
Ly$\alpha$ profile (solid black) that matches the observed spectrum (red). 
A best-fit is obtained for 
a WD mass of $1.03 M_{\odot}$ accreting at a rate of $7.6 \times 10^{-9}M_{\odot}$/yr,
with an inclination of $41^{\circ}$, giving a distance of 839~pc and 
bringing the $\chi^2_{\nu}$ value down to 2.30.  
\label{veileddisk} 
}
\end{figure}

From the (veiled) disk model we ran, we find that the best
fit is obtained for the larger WD mass ($1.03M_{\odot}$) 
and the lower inclination ($i=41^{\circ}$).
This model has $\chi^2_{\nu}=2.30$, and is presented in Fig.\ref{veileddisk}. 
While this model fit has the lowest $\chi^2_{\nu}$, it is only marginally
better than the other veiled disk models.   
We are unable to account for the broad Si+O absorption feature at $\sim 1300$~\AA ,
which can form in the WD photosphere (which gives an unrealistic distance) 
or in a thicker absorbing slab (which produces far more absorption lines 
than observed). Since the continuum is relatively well fitted, we are confident
that the mass accretion rate is of the order of $10^{-8}M_{\odot}$/yr 
for the $0.7 \pm 0.2 M_{\odot}$ WD mass and distance of $\approx 840$~pc.

\clearpage 

\section{Discussion and Conclusions} 

\par T Aur is clearly a disk dominated system. Preliminary calculations demonstrate that a lone white 
dwarf produces a luminosity far below the observed luminosity. While an exposed white dwarf fits 
the observed spectra (figure 2), the allowed distances are far below the Gaia distance. Evidence of 
the accretion disk is apparent in light curve analysis of T Aur. Both \cite{walker} and
\cite{bianchini} demonstrate small, rapid light variations which are evidence of an accretion disk.  

\par T Aur is an eclipsing binary. If the white dwarf is being eclipsed then the inclination of the 
system must be high. However, lower inclinations are allowed if a portion of the accretion disk is 
eclipsed, not the white dwarf. If the accretion hot spot is being eclipsed, not the white dwarf, then 
the inclination of 57\textsuperscript{o} reported by both \cite{sel19} 
and \cite{bianchini} would be consistent. \cite{walker} observe that the T Aur light curve has no 
secondary eclipse and that the eclipse depth varies from 0.10 to 0.28 mags. These variations in eclipse 
depth are evidence of an accretion hot spot eclipse. This idea was also expressed by \cite{bianchini}.

\par All accretion disk models fitted with absorbing slabs and ISM absorption have similar 
$\chi^2_{\nu}$ values. The model adopting a white dwarf mass of 1.03 $M_{\odot}$ and an inclination 
of 41\textsuperscript{o} produces the minimum $\chi^2_{\nu}$ value of 2.30. However, the model 
adopting a white dwarf mass of 0.8 $M_{\odot}$ and an inclination of 60\textsuperscript{o} produces 
a $\chi^2_{\nu}$ value of 2.55, which is not a statistically significant difference from the minimum. 
This model also produces a white dwarf mass within the allowed error range presented by
\cite{sel19} of 0.7$\pm$0.2 $M_{\odot}$. The inclination of 60\textsuperscript{o} is close 
to the value of 57\textsuperscript{o} reported by both \cite{sel19} and \cite{bianchini}. 
These qualifications lend merit to this particular model.

\section{Summary} 

\par T Aur is a disk-dominated old nova. Our calculations demonstrate the need for an 
accretion disk to account for the luminosity output of the system. The flickering present in 
the light curve and model fitting also confirm the disk. From model fitting, we determine a 
white dwarf mass of $0.80\:M_{\odot}$, a system inclination of $60^{\circ}$ and a mass 
transfer rate of $1.8 \times 10^{-8}M_{\odot}/$yr. We have compared our model fitting results to the FUV
data exclusively, with the results of Selvelli and Gilmozzi. Our disk model fitting for a
 $1.03 M_{\odot}$ white dwarf and inclination of 60 degrees
is also within their error bars of the accretion rate for a 1 solar mass white dwarf and an inclination of 
60 degrees. The relatively low inclination reveals the nature of the eclipses present in T Aur. In order to have a low 
inclination, part of the accretion disk, perhaps the hot spot, must be eclipsing. The white dwarf, being on the
order of planetary dimensions, cannot be eclipsed at an inclination of 60\textsuperscript{o}. While system parameters 
were determined, one mystery remains for T Aur.  The deep absorption lines of metals remain unexplained.
We know the the absorption lines do not form in the disk and they are too broad and deep to be interstellar. 
They are almost certainly associated with a circumbinary disk or possibly form in the shell 
ejected by the 1891 explosion. Because of the absence of subexposures (the STIS spectrum was a mere 600~s snapshot) 
and phase-resolved spectra,  new FUV spectra are needed for a full analysis of this eclipsing system. A Hubble COS spectrum in TIMETAG
mode down to the Lyman Limit would enable the detection of any motion of the absorption features.

\acknowledgements
CL acknowledges support by the Delaware Space Grant College and
Fellowship Program (NASA Grant 80NSSC20M0045). This research was also supported by HST Archival 
Grant AR-16152.001-A and NASA ADAP Grant 80NSSC21K0629, both to Villanova University. 
PG is very pleased to thank William P. Blair at the Henry A. Rowland
Department of Physics and Astronomy, Johns Hopkins University, Baltimore, Maryland, 
for his unabatingly kind hospitality.  

\software{
IRAF \citep[v2.16.1,][]{tod93}, 
Tlusty (v203) Synspec (v48) Rotin(v4) Disksyn (v7) \citep{hub17a,hub17b,hub17c}, 
Circus (v1) \citep{hub96}, PGPLOT (v5.2), Cygwin-X (Cygwin v1.7.16),
xmgrace (Grace v2), XV (v3.10) } 
\\ \\ 
 
\noindent  
Edward M. Sion \url{https://orcid.org/0000-0003-4440-0551} \\  
Patrick Godon \url{https://orcid.org/0000-0002-4806-5319}

\end{document}